\documentclass[%
 aip,
 amsmath,amssymb,
 reprint,%
]{revtex4-1}

\usepackage{graphicx}% Include figure files
\usepackage{dcolumn}% Align table columns on decimal point
\usepackage{bm}% bold math
\usepackage{xcolor}

\usepackage[utf8]{inputenc}
\usepackage[T1]{fontenc}
\usepackage{mathptmx}

\newcommand{\LMO}{Li$_2$MoO$_4$}
\newcommand{\onbb}{$0\nu2\beta$}

\DeclareUnicodeCharacter{2212}{-}
\begin{document}

% Use the \preprint command to place your local institutional report number 
% on the title page in preprint mode.
% Multiple \preprint commands are allowed.
%\preprint{}

\title{Phonon-mediated crystal detectors with metallic film coating capable of rejecting $\alpha$ and $\beta$ events induced by surface radioactivity} %Title of paper

% repeat the \author .. \affiliation  etc. as needed
% \email, \thanks, \homepage, \altaffiliation all apply to the current author.
% Explanatory text should go in the []'s, 
% actual e-mail address or url should go in the {}'s for \email and \homepage.
% Please use the appropriate macro for the type of information

% \affiliation command applies to all authors since the last \affiliation command. 
% The \affiliation command should follow the other information.

\author{I.C.~Bandac}
   \affiliation{Laboratorio Subterr\'aneo de Canfranc, 22880 Canfranc-Estaci\'on, Spain}
\author{A.S.~Barabash}
   \affiliation{National Research Centre Kurchatov Institute, Institute of Theoretical and Experimental Physics, 117218 Moscow, Russia\looseness=-1}
\author{L.~Berg\'e}
\author{Ch.~Bourgeois}
   \affiliation{Universit\'e Paris-Saclay, CNRS/IN2P3, IJCLab, 91405 Orsay, France}
\author{J.M.~Calvo-Mozota}
   \affiliation{Laboratorio Subterr\'aneo de Canfranc, 22880 Canfranc-Estaci\'on, Spain}
\author{P.~Carniti}
   \affiliation{INFN, Sezione di Milano Bicocca, I-20126 Milano, Italy}
\author{M.~Chapellier}
   \affiliation{Universit\'e Paris-Saclay, CNRS/IN2P3, IJCLab, 91405 Orsay, France}
\author{M. de Combarieu}
   \affiliation{IRFU, CEA, Universit\'e Paris-Saclay, F-91191 Gif-sur-Yvette, France}
\author{I.~Dafinei}
   \affiliation{INFN, Sezione di Roma, I-00185, Rome, Italy}
\author{F.A.~Danevich}
   \affiliation{Institute for Nuclear Research of NASU, 03028 Kyiv, Ukraine}
\author{L.~Dumoulin}
   \affiliation{Universit\'e Paris-Saclay, CNRS/IN2P3, IJCLab, 91405 Orsay, France}
\author{F.~Ferri}
   \affiliation{IRFU, CEA, Universit\'e Paris-Saclay, F-91191 Gif-sur-Yvette, France}
\author{A.~Giuliani}
\email{Andrea.Giuliani@ijclab.in2p3.fr.}
   \affiliation{Universit\'e Paris-Saclay, CNRS/IN2P3, IJCLab, 91405 Orsay, France}
\author{C.~Gotti}
   \affiliation{INFN, Sezione di Milano Bicocca, I-20126 Milano, Italy}
\author{Ph.~Gras}
   \affiliation{IRFU, CEA, Universit\'e Paris-Saclay, F-91191 Gif-sur-Yvette, France}
\author{E.~Guerard}
   \affiliation{Universit\'e Paris-Saclay, CNRS/IN2P3, IJCLab, 91405 Orsay, France}
\author{A.~Ianni}
   \affiliation{INFN, Laboratori Nazionali del Gran Sasso, I-67100 Assergi (AQ), Italy}
\author{H.~Khalife}
   \affiliation{Universit\'e Paris-Saclay, CNRS/IN2P3, IJCLab, 91405 Orsay, France}
\author{S.I.~Konovalov}
   \affiliation{National Research Centre Kurchatov Institute, Institute of Theoretical and Experimental Physics, 117218 Moscow, Russia\looseness=-1}
\author{P.~Loaiza}
\author{M. Madhukuttan}
\author{P.~de~Marcillac}
\author{R.~Mariam}
\author{S.~Marnieros}
\author{C.A.~Marrache-Kikuchi}
   \affiliation{Universit\'e Paris-Saclay, CNRS/IN2P3, IJCLab, 91405 Orsay, France}
\author{M.~Martinez}
\affiliation{Fundación ARAID \& Centro de Astropartículas y Física de Altas Energías, Universidad de Zaragoza, Zaragoza 50009, Spain\looseness=-1}
\author{C.~Nones}
   \affiliation{IRFU, CEA, Universit\'e Paris-Saclay, F-91191 Gif-sur-Yvette, France}
\author{E.~Olivieri}
   \affiliation{Universit\'e Paris-Saclay, CNRS/IN2P3, IJCLab, 91405 Orsay, France}
\author{G.~Pessina}
   \affiliation{INFN, Sezione di Milano Bicocca, I-20126 Milano, Italy}
\author{D.V.~Poda}
\author{Th.~Redon}
\author{J.-A.~Scarpaci}
   \affiliation{Universit\'e Paris-Saclay, CNRS/IN2P3, IJCLab, 91405 Orsay, France}
\author{V.I.~Tretyak}
   \affiliation{Institute for Nuclear Research of NASU, 03028 Kyiv, Ukraine}
\author{V.I.~Umatov}
   \affiliation{National Research Centre Kurchatov Institute, Institute of Theoretical and Experimental Physics, 117218 Moscow, Russia\looseness=-1}
\author{M.M.~Zarytskyy}
   \affiliation{Institute for Nuclear Research of NASU, 03028 Kyiv, Ukraine}
\author{A.S.~Zolotarova}
   \affiliation{Universit\'e Paris-Saclay, CNRS/IN2P3, IJCLab, 91405 Orsay, France}

%\author{A. Author}
% \altaffiliation[Also at ]{Physics Department, XYZ University.}%Lines break automatically or can be forced with \\
%\author{B. Author}%
 %\email{Second.Author@institution.edu.}
%\affiliation{ 
%Authors' institution and/or address%\\This line break forced with \textbackslash\textbackslash
%}%

%\author{C. Author}
 %\homepage{http://www.Second.institution.edu/~Charlie.Author.}
%\affiliation{%
%Second institution and/or address%\\This line break forced% with \\
%}%

%\author{}
%\email[]{Your e-mail address}
%\homepage[]{Your web page}
%\thanks{}
%\altaffiliation{}
%\affiliation{}

% Collaboration name, if desired (requires use of superscriptaddress option in \documentclass). 
% \noaffiliation is required (may also be used with the \author command).
%\collaboration{}
%\noaffiliation

\date{\today}

\begin{abstract}
% insert abstract here
Phonon-mediated particle detectors based on single crystals and operated at millikelvin temperatures are used in rare-event experiments for neutrino physics and dark-matter searches. In general, these devices are not sensitive to the particle impact point, especially if the detection is mediated by thermal phonons. In this letter, we demonstrate that excellent discrimination between interior and surface $\beta$ and $\alpha$ events can be achieved by coating a crystal face with a thin metallic film, either continuous or in the form of a grid. The coating affects the phonon energy down-conversion cascade that follows the particle interaction, leading to a modified signal shape for close-to-film events. An efficient identification of surface events was demonstrated with detectors based on a rectangular $20 \times 20 \times 10$~mm$^3$ \LMO~crystal coated with a Pd normal-metal film (10~nm thick) and with Al-Pd superconductive bi-layers (100~nm-10~nm thick) on a $20 \times 20$~mm$^2$ face. Discrimination capabilities were tested with $^{238}$U sources emitting both $\alpha$ and $\beta$ particles. Surface events are identified for energy depositions down to millimeter-scale depths from the coated surface. With this technology, a substantial reduction of the background level can be achieved in experiments searching for neutrinoless double-beta decay. %This work is supported by the European Commission (Project CROSS, Grant ERC-2016-ADG, ID 742345).
\end{abstract}

% insert suggested PACS numbers in braces on next line

\maketitle %\maketitle must follow title, authors, abstract and \pacs

% Body of paper goes here. Use proper sectioning commands. 
% References should be done using the \cite, \ref, and \label commands
%\section{}
%\label{}
%\subsection{}
%\subsubsection{}

Phonon-mediated particle detectors~\cite{enss:2008a} (often defined ``bolometers'') have nowadays important applications in neutrino physics,~\cite{giuliani:2012a,nucciotti:2014a} dark-matter searches~\cite{pirro:2017a} and rare nuclear decay investigations.~\cite{belli:2019a} They also provide outstanding $\alpha$, $\beta$, $\gamma$, X-ray and neutron spectroscopy.~\cite{enss:2008a,pirro:2017a,belli:2019a,bekker:2016a} 

Neutrinoless double-beta (\onbb) decay~\cite{dolinski:2019a} is a hypothetical rare nuclear transition of an even-even nucleus to an isobar with two more protons, with the emission of just two electrons. Its observation would provide a unique insight into neutrino physics.~\cite{vergados:2016a} Bolometers based on \LMO~crystals are promising detectors for a next-generation \onbb~decay experiment.~\cite{bekker:2016a,armengaud:2017a,armengaud:2020a} They embed the favorable candidate $^{100}$Mo, maximising the detection efficiency. The \onbb~decay signature is a peak in the sum energy spectrum of the two emitted electrons, expected at 3.034~MeV for $^{100}$Mo. In a bolometer, the energy deposited by a particle in the crystal is converted into phonons, which are then detected by a suitable sensor. 

The highest challenge in \onbb~decay search is the control of the radioactive background, due to the long expected lifetime of the process ($> 10^{25}-10^{26}$~y).~\cite{gando:2016a,agostini:2020a,adams:2019a} The experiments are located underground under heavy shielding. To reduce the current background level of bolometric experiments, it is mandatory to reject $\alpha$ or $\beta$ events --- defined ``surface $\alpha$'s or $\beta$'s'' in the following for brevity --- induced by radioactive impurities located either close to the surface of the crystal itself or to that of the surrounding structure.\cite{artusa:2014a,alduino:2017a} Surface $\alpha$'s can be rejected in scintillating materials --- such as \LMO~--- by detecting simultaneously scintillation and phonons for the same event~\cite{pirro:2006a,artusa:2014a,poda:2017a} and exploiting the generally lower light yield of $\alpha$'s with respect to $\beta$'s,~\cite{tretyak:2010a} but the rejection of surface $\beta$'s requires dedicated techniques capable of tagging surface events in bolometers.~\cite{foggetta:2005a,marnieros:2008a,nones:2010a,nones:2012a,agnese:2013a}

In this letter, we report an effective method to identify both surface $\alpha$'s and $\beta$'s in \LMO~bolometers. The discrimination is achieved by coating a \LMO~crystal side with a metallic film acting as a pulse-shape modifier for events that release energy close to the coated face. When an ionizing event occurs in a dielectric crystal kept at mK temperature, the deposited energy is readily converted to athermal phonons with typical energies of the order of tens of meV, to be compared with the few $\mu$eV thermal-bath energy.
The energy down-conversion of these athermal phonons occurs mainly by anharmonic decay and is progressively slowing down, as the phonon lifetime scales as the fifth power of the energy.~\cite{orbach:1964a,bron:1982a} If a sensor sensitive mainly to thermal phonons is used (as in this work), the rise time of the signal is in the $\sim 10$~ms range, which corresponds to the typical thermalization time of the deposited energy. However, thermalization can speed up via a metallic film covering a crystal side. If the particle is absorbed close to the film, a significant fraction of its energy is trapped in the metal in the form of hot electrons, excited by the absorption of the particle-generated athermal phonons. The energy is quickly thermalised in the electron system, so that phonons of much lower energies are re-injected in the crystal from the film. Signals from events occurring close to the film will present therefore a shorter rise time and a modified time evolution. We show here that surface events can be tagged according to this approach.  

All the detectors in this work share a common basic structure, which is similar to that used in the \onbb~experiments CUORE,~\cite{adams:2019a} LUMINEU,~\cite{armengaud:2017a} CUPID-Mo,~\cite{armengaud:2020a,armengaud:2020b} CUPID-0~\cite{azzolini:2019a} and in the dark-matter experiment EDELWEISS~\cite{armengaud:2017b} as far as the phonon readout is concerned. The surface sensitivity was studied above ground with prototypes of reduced size with respect to the final \onbb~bolometers. The energy absorber of the bolometers described here is a single \LMO~crystal~\cite{grigorieva:2017a} with a size of $20 \times 20 \times 10$~mm$^3$ and a mass of $\sim 12$~g. All the tests involve just a single $20 \times 20$~mm$^2$ coated side. 

\begin{figure}[t]
\includegraphics[scale=0.26]{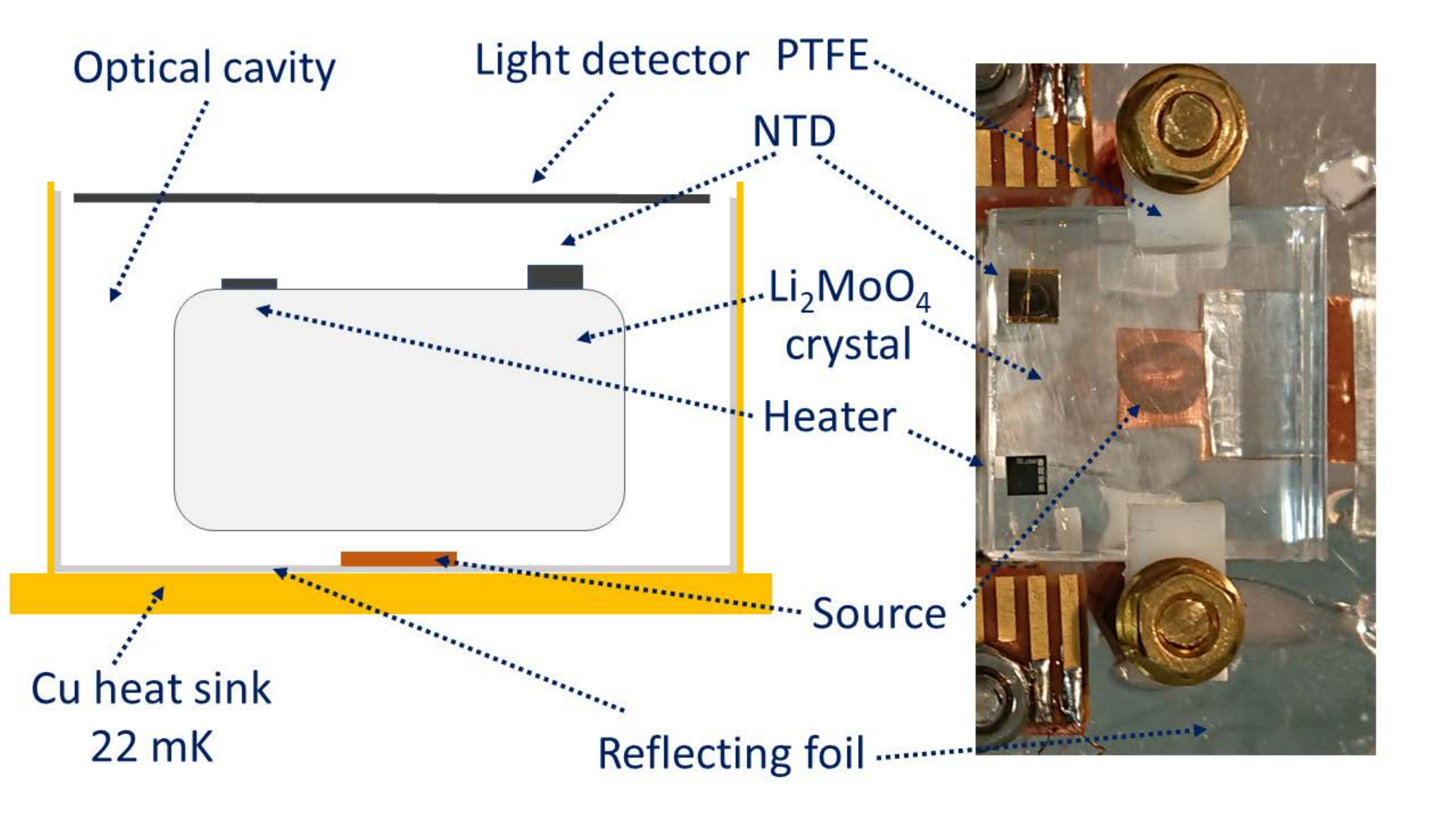}
\caption{\label{fig:detector-assembly} Scheme (left) and photograph (right) of the detector assembly with a bare \LMO~crystal. A Ge thermistor and a Si:P heater (used to stabilize the bolometric response) are glued on the upper face of the crystal, which is held by polytetrafluoroethylene (PTFE) elements, not shown in the scheme. A uranium source --- visible in transparency in the photograph --- is placed below the crystal. A bolometric light detector --- removed to take the photograph --- faces the upper side of the crystal. The reflecting foil forms an optical cavity that aids light collection.}
\end{figure}

The phonon sensor is a neutron transmutation doped Ge thermistor~\cite{haller:1984a} (NTD) with a size of $3 \times 3 \times 1$~mm$^3$. Its resistivity increases exponentially as the temperature decreases.~\cite{efros:1975a} The NTD is glued on the crystal by means a two-component epoxy. The glue provides a slow transmission interface, making the NTD sensitive mainly to thermal phonons. 

We used uranium radioactive sources to test the detector surface sensitivity. They were obtained by drying up a drop of uranium acid solution on a copper foil. These sources provide two main $\alpha$ lines at $\sim 4.2$ and $\sim 4.7$~MeV from $^{238}$U and $^{234}$U respectively, affected by a significant straggling due to partial $\alpha$ absorption in the source residues and/or in the copper substrate. $^{238}$U disintegration is followed by two consecutive $\beta$ emissions, from $^{234}$Th (with a half-life of 24.1~d and an end-point of 0.27~MeV) and from $^{234m}$Pa (with a half-life of 1.2~min and an end-point of 2.27~MeV). The $^{238}$U $\alpha$ rate and the $^{234m}$Pa $\beta$ rate are extremely close.

\begin{figure*}[t]
\includegraphics[scale=0.48]{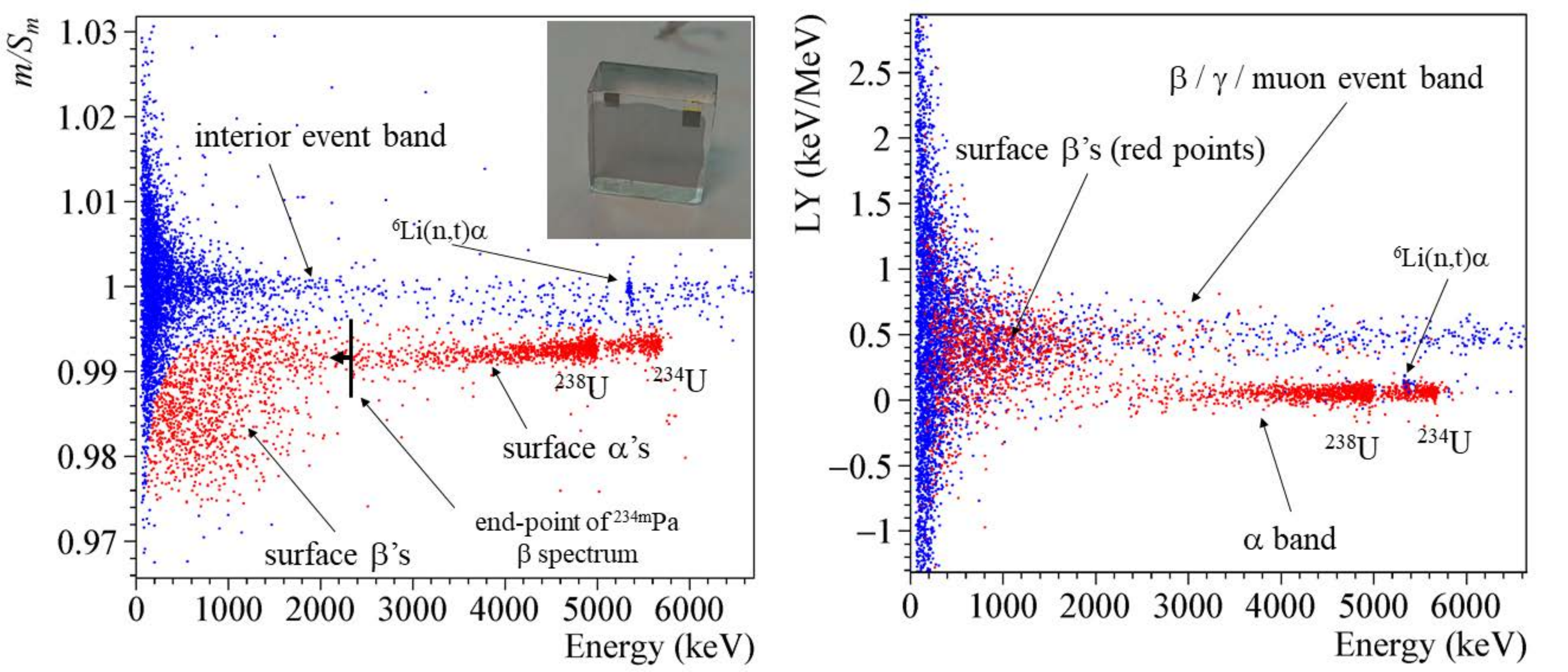}
\caption{\label{fig:Pd-film} Particle identification obtained by a \LMO~detector with a 10-nm-thick Pd coating (see inset on the left) exposed to a uranium source. Left: The pulse-shape parameter $m/S_m$ is plotted as a function of the heat energy (estimated by $S_m$) deposited in the \LMO~crystal, with a $\gamma$-based calibration. Surface events appear as a population with lower $m/S_m$ values. They are selected by visual inspection and highlighted in red. The neutron-capture line from the reaction $^6$Li(n,t)$\alpha$ lays in the interior-event band. The $\alpha$ particles are mis-calibrated by about $20$~\% due to both pulse-shape effects and intrinsic different responses for $\alpha$'s and $\beta$/$\gamma$'s. Right: The \LMO~scintillation light yield (LY) is plotted for the same pulses on the same energy scale. The LY is expressed by the ratio of the energy deposited in the light detector by scintillation photons (in keV) to that deposited in the \LMO~crystal as heat (in MeV), for the same event. The same surface events highlighted in the left panel are shown in red. Surface $\beta$'s lay in the high-LY band, while $\alpha$'s and the neutron capture events are well separated in the low-LY band.
}
\end{figure*}

The detector assembly is shown in Fig.~\ref{fig:detector-assembly}. A first test was conducted with a \LMO~crystal without coating to establish the bare detector performance. All subsequent tests have adopted the configuration shown in Fig.~\ref{fig:detector-assembly}, where the metal-coated side, which is optically polished before the film deposition, faces the radioactive source. A bolometric light detector based on a Ge wafer~\cite{armengaud:2017a,armengaud:2020a} is used to separate $\alpha$ from $\beta$/$\gamma$/muon events by detecting the scintillation light.~\cite{pirro:2006a,artusa:2014a,poda:2017a}

The detectors were cooled down in a dilution refrigerator located in IJCLab, Orsay, France.~\cite{mancuso:2014a} All the data discussed here have been collected with the detector thermalized to a copper plate at $\sim 22$~mK (Fig.~\ref{fig:detector-assembly}). A current of the order of $\sim 5$~nA is injected in the NTD, rising the detector temperature to about $\sim 25$~mK, at which the NTD resistance is about 0.5~M$\Omega$. The voltage signal amplitude across the NTD is $\sim 60$~$\mu$V/MeV for the bare crystal, corresponding to an NTD temperature change of $\sim 0.5$~mK/MeV. The pulse rise time (from 10\% to 90\% of maximum amplitude) is typically in the 3--10~ms range and the pulse decay time (from 90\% to 30\% of maximum amplitude) is tens of ms. The signals are read out by a DC-coupled low-noise voltage-sensitive amplifier.~\cite{arnaboldi:2002a} In all the tests, the \LMO~detector is energy-calibrated using $\gamma$~peaks of the environmental background, and the light detector using the cosmic muons crossing the Ge wafer~\cite{novati:2019a}. In the \LMO~heat channel, we obtained routinely good energy resolutions of 5--10~keV FWHM for environmental $\gamma$ peaks in the 0.2--1~MeV region.

The first test to attain surface sensitivity was performed with a 10-$\mu$m-thick Al-film coating. The details of the achieved results are reported elsewhere.~\cite{bandac:2020a,khalife:2020a} We remind here that an excellent separation of surface $\alpha$ particles was demonstrated thanks to pulse-shape discrimination.

The best separation between surface $\alpha$'s and any type of interior events was obtained via a specially developed pulse-shape parameter --- extensively used here --- that we will designate as $m/S_m$.~\cite{bandac:2020a} To construct it, the signals are passed through a digital optimal filter,~\cite{gatti:1986a} whose transfer function is built using the noise power spectrum and the pulse shape of an interior event. This filter provides the best estimator of the signal amplitude $S_m$ (i.e. energy). An individual pulse $S(t)$ is plotted point by point against an average pulse $A(t)$ --- formed from a large sample of interior events and normalized to~1 --- obtaining approximately a straight line. The related slope parameter $m$ is an estimator of the pulse amplitude as well. The ratio $m/S_m$ turns out to be very sensitive to the pulse shape. Interior events have $m/S_m \sim 1$, as expected. On the contrary, $m/S_m$ deviates from~1 for surface $\alpha$ events. 

For the Al-coated \LMO~crystal, the separation between the interior and the surface $\alpha$ events from a uranium source is better than $10 \sigma$ in terms of $m/S_m$ distributions.~\cite{bandac:2020a} Unfortunately, only a slight hint of separation of the surface $\beta$ events emitted by the same source was observed,~\cite{bandac:2020a,khalife:2021a} ruling out Al coating as a viable method for a complete surface-event tagging.

Aluminum was chosen as it is superconductive at the bolometer operation temperature, with a critical temperature $T_C (\mathrm{Al}) \sim 1.2$~K.~\cite{Cochran:1958a} This leads to a negligible contribution to the heat capacity of the full bolometer, as the electron specific heat of superconductors vanishes exponentially with the temperature. We remark that the heat capacity of a bolometer must be as low as possible to achieve high signal amplitudes. In fact, no deterioration of the detector sensitivity was observed with respect to the bare \LMO~crystal. However, the behaviour of superconductors can spoil surface particle tagging. The prompt absorption of athermal phonons by the film breaks Cooper pairs and forms quasi-particles. 
Theoretically, quasi-particle lifetime diverges as the temperature of the superconductor decreases,~\cite{kaplan:1976a,barends:2008a} although it is often experimentally found to saturate at low temperatures.~\cite{devisser:2014a,fyhrie:2020a} In aluminum, at very low temperatures such as ours ($T/T_C < 0.02$), we expect the quasi-particle lifetime to be as large as several ms,~\cite{schnagel:2000a,baselmans:2009a,fyhrie:2020a,devisser:2014a} similar to the thermalization time of interior events. 
This mechanism competes with the faster thermalization that should be provided by the film. 

Driven by these considerations, we tested a \LMO~bolometer with a normal-metal coating. At low temperatures, the electron specific heat of normal metals is proportional to the temperature and tends to dominate over the crystal heat capacity, which scales as $T^3$ according to the Debye law. The thickness of normal-metal films must be substantially smaller than the aluminum ones. We chose palladium as a coating material as it can provide continuous thin films down to 2~nm thickness and no challenging radioactive isotopes are present in its natural composition. A thickness of 10~nm was chosen as a good compromise between heat capacity reduction and phonon absorption probability. The particle-identification results are encouraging, as shown in Fig.~\ref{fig:Pd-film}: both surface $\alpha$'s and $\beta$'s are well separated from the interior events. 

Unfortunately, the heat capacity of the Pd film~\cite{mizutani:2001a} competes with that of the \LMO~crystal~\cite{musikhin:2015a} affecting seriously the sensitivity of the detector, which was only $\sim 23$~$\mu$V/MeV, about one third of that achieved with the bare crystal. Therefore, this option is not viable for a full coating of the crystal.

\begin{figure*}[t]
\includegraphics[scale=0.49]{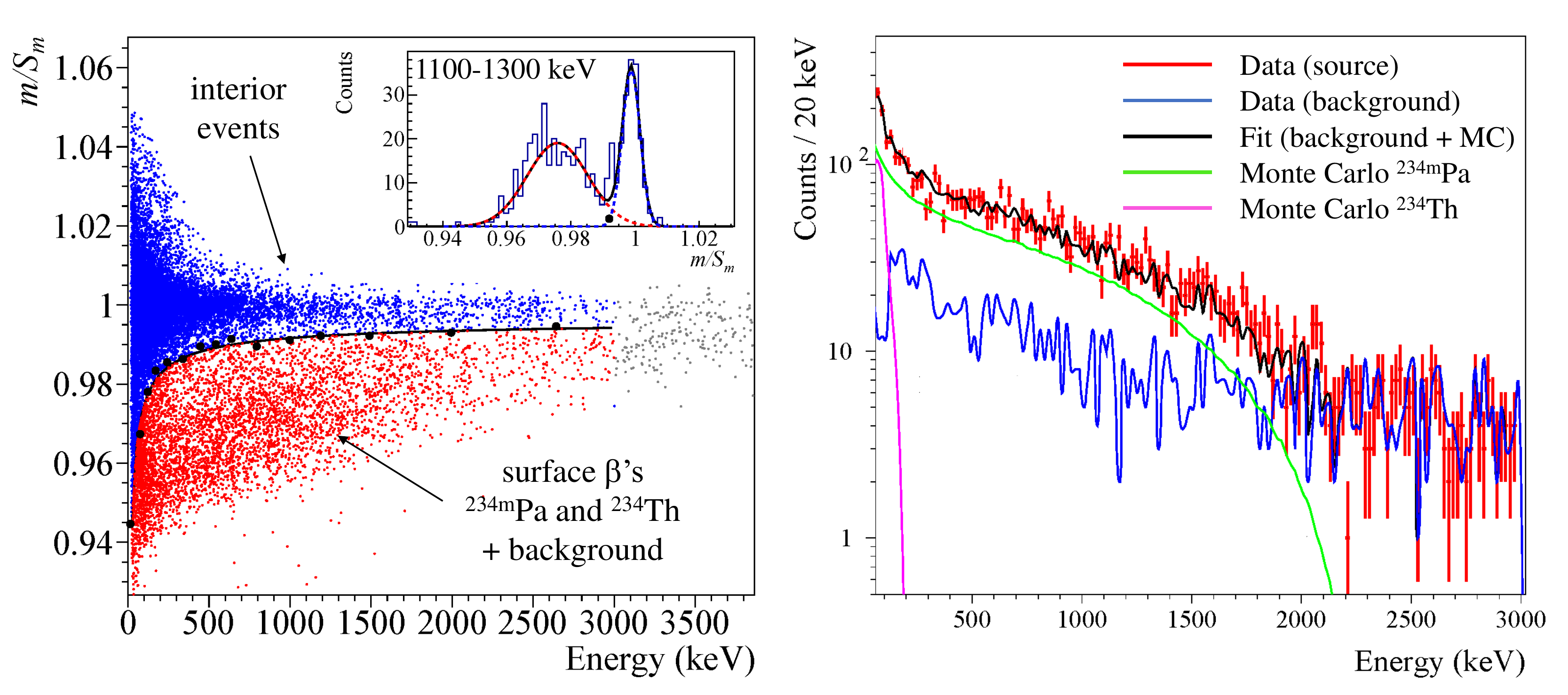}
\caption{\label{fig:Al-Pd} Particle identification obtained by a \LMO~detector with an Al-Pd coating exposed to a uranium source. The $\alpha$ events are removed by a light-yield cut. Left: in a plot of the pulse-shape parameter $m/S_m$ versus energy, the surface events (in red) lay below a black curve defining a 3$\sigma$ acceptance region for the interior events (in blue). The analysis is carried out in the [0-3000]~keV energy interval, which is divided in several sub-intervals. For each of them, a double Gaussian fit of the $m/S_m$ distribution is performed to separate the two populations. An example is provided in the inset. The black point is located 3$\sigma$ at the left of the mean of the Gaussian of the interior events. The black curve fits the black points by a power-law function. Right: Energy spectra (with and without source) of the surface events selected according to the procedure illustrated on the left. The live times of the two measurements are normalized. The fit of the source data accounts for the two simulated $\beta$ contributions of the uranium source and that of the background. 
}
\end{figure*}

To overcome the heat-capacity problem, we developed a detector coated with an Al-Pd bi-layer (100~nm and 10~nm thick respectively, with Al on the top), which is superconducting by proximity effect below $T_C$(Al-Pd)~$= 0.65$~K. The superconductive gap induced in Pd by the Al film reduces substantially the Pd specific heat with respect to the normal state. This gap is however low enough to ensure the fast thermalization of the energy deposited by surface events. In fact, the surface-event discrimination capability was fully maintained (see Fig.~\ref{fig:Al-Pd}, left). The detector sensitivity was measured to be 43~$\mu$V/MeV, almost doubled with respect to the pure Pd film. 

\begin{figure}[t]
\includegraphics[scale=0.49]{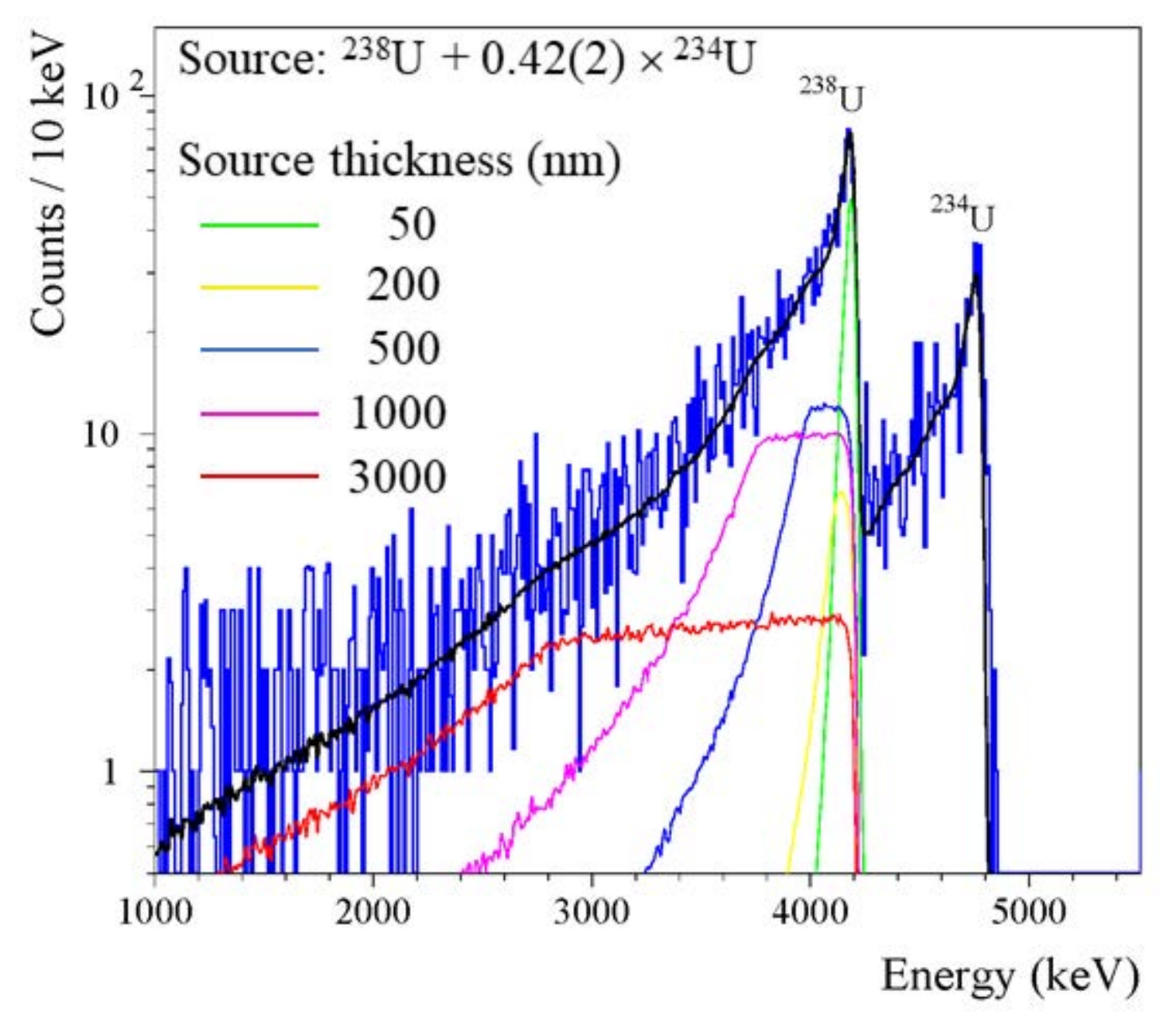}
\caption{\label{fig:alpha-source} Energy spectrum collected by a \LMO~detector with an Al-Pd coating exposed to a uranium source after selection of the $\alpha$ events by a light-yield cut with $\sim 100$\% efficiency. The spectrum is calibrated using the $\alpha$-line positions. The measurement is the same that provided the source data shown in Fig.~\ref{fig:Al-Pd}. The straggling can be reproduced by assuming five source components in copper. Each component is a 6~mm diameter disk with a given thickness. The active nuclei are assumed to be uniformly distributed in each disk. The exact source structure is unknown, but our goal here is to set up a phenomenological model capable of explaining the observed straggling. $^{238}$U and $^{234}$U are not in secular equilibrium, as already observed in these types of liquid sources.}
\end{figure}

\begin{figure}[t]
\includegraphics[scale=0.46]{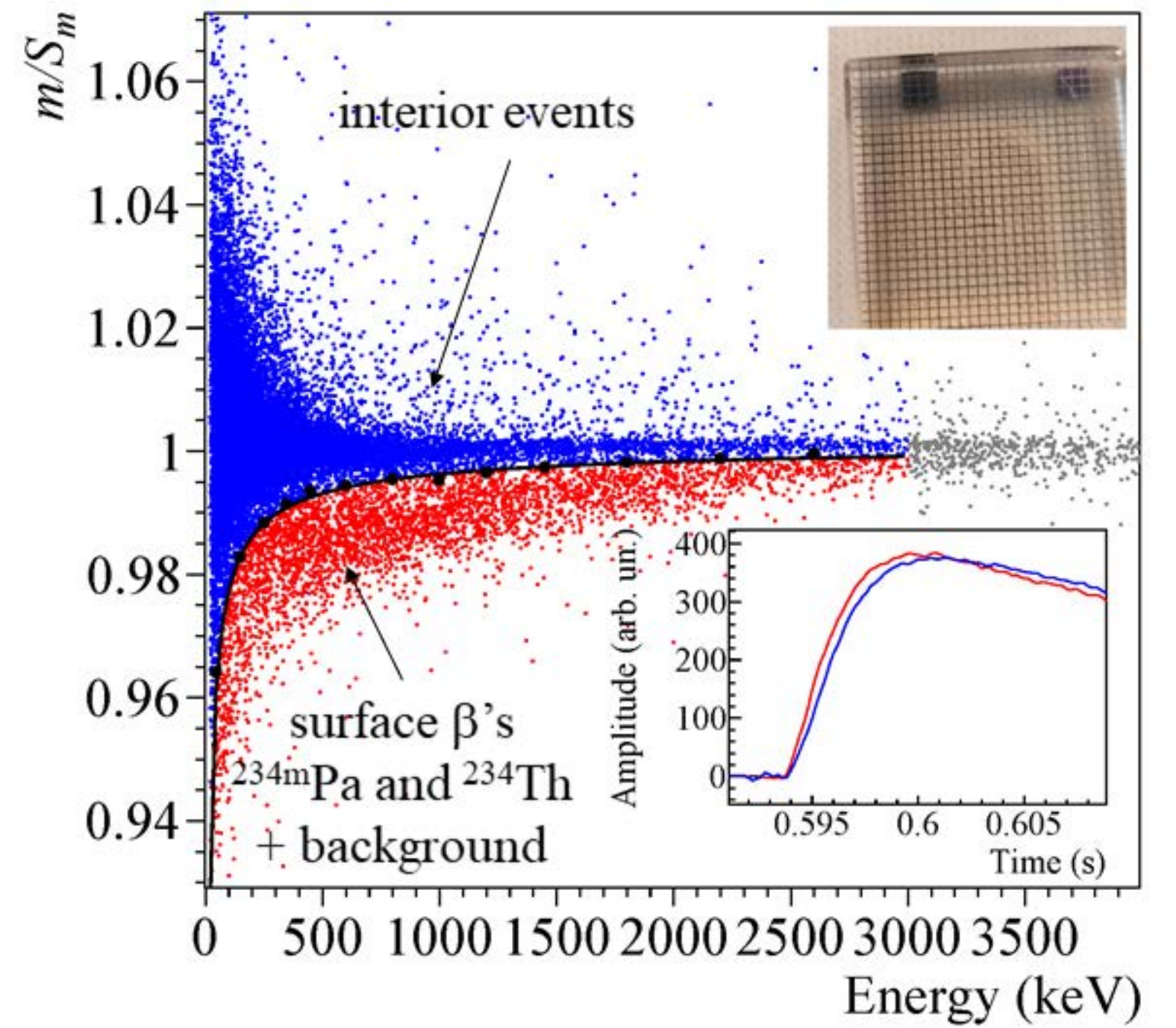}
\caption{\label{fig:grid} Particle identification obtained by a \LMO~detector with an Al-Pd grid coating exposed to a uranium source. The event selection is performed as in Fig.~\ref{fig:Al-Pd}, left. In the top inset,  the grid-coated crystal is shown. In the bottom inset, pulses from a surface (red) and an interior (blue) event are shown, corresponding to a deposited energy of about 1~MeV.}
\end{figure}

We performed two runs with the bi-layer detector. In the first, a uranium source was present, while the second was a background measurement in the same configuration. The trigger rate was $\sim 0.2$~Hz with the source. The contribution of the source is at the level of $\sim 0.03$~Hz. First, we developed a method to separate the surface $\beta$ component. The events below the black curve in the left panel of Fig.~\ref{fig:Al-Pd} --- collected in a source run --- are selected as surface events, while those above represent more than 99\% of the interior event population. The same analysis was performed for the background run. By means of Geant4-based~\cite{agostinelli:2003a} Monte-Carlo simulations (using the G4RadioactiveDecay and Decay0~\cite{ponkratenko:2000a} event generators), we were then able to confirm that the surface $\beta$ events isolated at low energies come actually from the radioactive source. We built a model to predict the $\beta$ spectrum shape considering the observed $\alpha$ straggling (Fig.~\ref{fig:alpha-source}) and the $\beta$ interactions in the detector. We fitted then the experimental $\beta$ spectrum using the predicted shape and taking the background into account (right panel in Fig.~\ref{fig:Al-Pd}). The total number of $^{234m}$Pa decay events returned by the fit --- the only free parameter --- is 3526(81). To build the source model, we set a uranium-source depth profile capable of reproducing the observed $\alpha$ spectrum and the related straggling, as shown in Fig.~\ref{fig:alpha-source}. From the model and the experimental number of $\alpha$ counts it was possible to predict independently the expected total number of $^{234m}$Pa events, which resulted to be 3455(273), in excellent agreement with that deduced from the selection of the source $\beta$ events. The efficiency in selecting surface $\beta$ events can be estimated as 102(8)\%. 
 
The $\beta$-particle range in \LMO~is of the order of 2~mm at 1~MeV and 4~mm at 2~MeV. Therefore, we can separate events that deposit a significant amount of energy up to $\sim 4$~mm from the film, well beyond its thickness. We performed then a last test by replacing the continuous Al-Pd film with an Al-Pd grid. The width of the grid lines was 70~$\mu$m and the spacing between each line was 700~$\mu$m (see inset in Fig.~\ref{fig:grid}). The purpose of using a grid is manifold: (1) further reduction of the heat capacity of the coating; (2) possibility to extract scintillation light through the coating; (3) availability of geometrical parameters to possibly tune the discrimination depth. The grid was tested with another uranium source, prepared with the same method as the first one, but about twice less intense. 
The detector with grid coating can separate surface $\beta$ events (see Fig.~\ref{fig:grid}). The $\beta$ selection efficiency resulted to be 93(10)\%, in good agreement with the continuous-film results. In addition, we measured a discrimination power of about 4.5$\sigma$ for surface $\alpha$ events using the $m/S_m$ parameter. In terms of detector performance, we observed an almost full recovering of the detector sensitivity, that was $\sim 51$~$\mu$V/MeV for $\beta$/$\gamma$ events. Therefore, the grid method is currently our protocol for surface event discrimination. 

In conclusion, we have shown that both $\alpha$ and $\beta$ particles absorbed close to a metal-coated surface of a \LMO~bolometer can be rejected with high efficiency by pulse-shape discrimination. The prospects of this approach for \onbb~searches are promising. In fact, the current background model of the future \onbb~experiment CUPID~\cite{CUPID:2019a} predicts a background level of 0.1~counts/(tonne~y~keV). Next-to-next generation experiments aim to a reduction of an additional factor 10. Since surface $\beta$ events contribute significantly to the current background level, a necessary condition to achieve the desired reduction is to reject them with an efficiency up to 90\%. This is achievable with the technique here described. 

This work is supported by the European Commission (Project CROSS, Grant ERC-2016-ADG, ID 742345). The ITEP group was supported by the Russian Scientific Foundation (grant No. 18-12-00003). F.A.D., V.I.T. and M.M.T. were supported in part by the National Research Foundation of Ukraine (grant No. 2020.02/0011). The PhD fellowship of H.K. has been partially funded by the P2IO LabEx (ANR-10-LABX-0038) managed by the Agence Nationale de la Recherche (France). The dilution refrigerator used for the tests and installed at IJCLab (Orsay, France) was donated by the Dipartimento di Scienza e Alta Tecnologia of the Insubria University (Como, Italy).

The data that support the findings of this study are available from the corresponding author upon reasonable request.

\nocite{*}
\bibliography{CROSS-APL-short}% Produces the bibliography via BibTeX.

\end{document}